\documentclass[prd,aps,twocolumn,preprintnumbers,showpacs, nofootinbib,superscriptaddress,notitlepage]{revtex4-1}
\usepackage{amsmath,graphicx,epsfig,amssymb}
\usepackage{inputenc}
\usepackage[usenames]{color}
\usepackage{slashed}
\usepackage{multirow}
\usepackage{subfigure}
\usepackage{dsfont}

\allowdisplaybreaks

\usepackage{floatrow}
\newfloatcommand{capbtabbox}{table}[][\FBwidth]

\begin{document}

\title{Hybrid Renormalization with Gradient Flow for Baryon Quasi-Distribution Amplitudes} 

\author{Jia-Lu Zhang}
\email{elpsycongr00@sjtu.edu.cn}
\affiliation{Tsung-Dao Lee Institute, Shanghai Jiao Tong University, Shanghai 201210, China}
\affiliation{State Key Laboratory of Dark Matter Physics, School of Physics and Astronomy, Shanghai Jiao Tong University,  Shanghai 200240, China}

\author{Mu-Hua Zhang}
\affiliation{State Key Laboratory of Dark Matter Physics, School of Physics and Astronomy, Shanghai Jiao Tong University,  Shanghai 200240, China}
\affiliation{Tsung-Dao Lee Institute, Shanghai Jiao Tong University, Shanghai 201210, China}
\begin{abstract}

We establish a factorization relation between baryon quasi-distribution amplitudes (quasi-DAs) defined with gradient flow and their counterparts renormalized in the $\overline{\mathrm{MS}}$ scheme. Working beyond the small flow-time limit, we perform a complete one-loop calculation that yields the full matching kernel and the associated Wilson-line linear divergence for flowed quasi-DAs. Building on this relation, we formulate a hybrid renormalization scheme that combines ratio scheme with gradient-flow matching. The framework enables a systematic connection between lattice QCD matrix elements and continuum baryon quasi-DAs, with reduced renormalization uncertainties and clear guidance for practical lattice analyses.
\end{abstract}

\maketitle
\section{Introduction}
Understanding the partonic structure of hadrons from first principles remains a central goal of Quantum Chromodynamics (QCD). Among the key nonperturbative quantities in this pursuit are the light-cone distribution amplitudes (LCDAs), which encode how the longitudinal momentum of a hadron is shared among its valence quarks and gluons. LCDAs play a crucial role in QCD factorization theorems for hard exclusive processes, such as meson and baryon form factors at large momentum transfer~\cite{Lepage:1980fj,Chernyak:1983ej} and exclusive meson production~\cite{Mankiewicz:1997uy}. However, because they are defined through nonlocal light-cone correlations, LCDAs are intrinsically Minkowskian and cannot be accessed directly in Euclidean lattice QCD. Their determination therefore requires indirect or alternative approaches.

A powerful strategy is to construct quasi-distribution amplitudes (quasi-DAs) from equal-time correlators of boosted hadrons, which can be computed on the lattice and related to LCDAs through perturbative matching. This approach, realized in frameworks such as Large Momentum Effective Theory (LaMET)~\cite{Ji:2013dva,Ji:2014gla,Izubuchi:2018srq}, the pseudo-distribution method~\cite{Radyushkin:2017cyf,Radyushkin:2019owq} and other related methods~\cite{Ma:2014jla,Ma:2017pxb}, has enabled systematically improvable studies of LCDAs~\cite{Han:2023hgy,Han:2024ucv,Deng:2023csv,Han:2024cht,Han:2025odf,LatticeParton:2024vck,Han:2023xbl,Hu:2023bba,LatticeParton:2022zqc,Hua:2020gnw,Wang:2019msf,Wang:2024wwa,LatticeParton:2024zko,Han:2024fkr,Braun:2018brg} and other observables such as parton distribution functions (PDFs) and generalized parton distributions (GPDs)~\cite{LatticePartonLPC:2022eev,Guo:2025obm,Wang:2025uap,LatticeParton:2023xdl,LatticePartonCollaborationLPC:2022myp,Deng:2022gzi,LatticeParton:2020uhz,Wang:2019tgg,Zhang:2023bxs,Wang:2025uap,Xiong:2013bka,Lin:2014zya,Alexandrou:2015rja,Chen:2016utp,Alexandrou:2016jqi,Alexandrou:2018pbm,Chen:2018xof,Wang:2017qyg,Wang:2017eel,Lin:2018pvv,LatticeParton:2018gjr,Alexandrou:2018eet,Liu:2018hxv,Zhang:2018nsy,Izubuchi:2018srq,Izubuchi:2019lyk,Chen:2020arf,Chen:2020iqi,Chen:2020ody,Shugert:2020tgq,Chai:2020nxw,Lin:2020ssv,Fan:2020nzz,Gao:2021hxl,Gao:2021dbh,Gao:2022iex,Su:2022fiu,LatticeParton:2022xsd,Gao:2022uhg,Gao:2023ktu,Gao:2023lny,Chen:2024rgi,Holligan:2024umc,Holligan:2024wpv}. In recent years, this method has been successfully applied to baryon LCDAs both theoretically and numerically~\cite{Deng:2023csv,Han:2023xbl,Han:2023hgy,Han:2024ucv}.

Despite these advances, two major challenges persist in practical lattice implementations. The first is the poor signal-to-noise ratio in nonlocal correlation functions, which becomes particularly severe at large hadron momenta and long Wilson-line separations. Reliable access to light-cone physics requires both large $P$ and sizable $z$, but increasing either worsens statistical precision. The second difficulty lies in renormalization: since continuum schemes such as $\overline{\mathrm{MS}}$ cannot be implemented directly on the lattice, one must employ an intermediate nonperturbative scheme, most commonly RI/MOM~\cite{Constantinou:2017sej,Stewart:2017tvs}. While effective in principle, it requires a precise gauge fixing to control systematic uncertainties~\cite{Zhang:2024omt}. Alternative approaches such as self-renormalization and hybrid renormalization~\cite{LatticePartonLPC:2021gpi,Ji:2020brr,Chou:2022drv} mitigate some systematics but are often more complicated to implement in practice.

The gradient flow formalism~\cite{Narayanan:2006rf,Luscher:2010iy,Luscher:2011bx,Luscher:2013cpa} offers a promising solution to these issues. By evolving gauge and fermion fields along a fictitious fifth (flow-time) dimension governed by diffusion-like equations, the flow acts as a gauge-invariant smearing over a physical length scale $\sqrt{8t}$, where $t$ is the flow time. It systematically suppresses ultraviolet (UV) fluctuations while preserving long-distance physics. As a result, most composite operators built from flowed fields are finite and well-defined at nonzero flow time, eliminating the need for additional renormalization and improving the numerical stability of correlation functions on the lattice.

Moreover, the gradient flow enhances the signal quality of nonlocal or boosted operators essential for quasi-distribution studies. The smearing reduces short-distance statistical noise and improves the overlap with the ground-state signal, leading to cleaner plateaus and more precise lattice measurements. In practice, this results in improved statistical precision and better control over systematic uncertainties. Despite the smearing, flowed operators admit a well-defined perturbative expansion and can be matched to their counterparts in the $\overline{\mathrm{MS}}$ scheme. These properties make the gradient flow a powerful tool for combining the nonperturbative strengths of lattice QCD with the analytical control of continuum perturbation theory.

In this work, we establish the factorization relation between quasi-DAs defined with gradient flow and their counterparts renormalized in the $\overline{\mathrm{MS}}$ scheme. A complete one-loop calculation beyond the small-flow-time approximation is performed, yielding the full matching kernel and the linear divergence of the flowed quasi-DAs. Based on this relation, we propose a hybrid renormalization scheme, in which gradient-flow matching is used to convert the lattice scheme to the $\overline{\mathrm{MS}}$ scheme, and the ratio scheme is employed to eliminate the logarithmic divergence. This framework provides a systematic and practical approach to connect lattice QCD calculations of baryon quasi-DAs with continuum observables, offering improved control over renormalization uncertainties.

The remainder of this paper is organized as follows. Section~II introduces the theoretical framework, including the definitions of baryon LCDAs and quasi-DAs and the gradient flow formalism. Section~III establishes the factorization relation between flowed quasi-DAs and their $\overline{\mathrm{MS}}$ counterparts and presents the complete one-loop calculation beyond the small-flow-time approximation. Section~IV develops the hybrid renormalization scheme. Section~V concludes with a brief summary and outlook.

\section{Theoretical Framework}

\subsection{Quasi-DAs and LCDAs for Baryons}

The baryon light-cone distribution amplitudes (LCDAs) are defined through nonlocal hadron-to-vacuum matrix elements with light-like separations~\cite{Braun:1999te}:
\begin{equation}
\begin{aligned}
{H(z_1,z_2,z_3)}_{\alpha\beta\gamma} 
& = \epsilon^{ijk} \langle 0 | 
f_{\alpha}^{i'}(z_1 n) W^{i'i}(z_1 n, z_0 n) \\
& \times g_{\beta}^{j'}(z_2 n) W^{j'j}(z_2 n, z_0 n) \\
& \times h_{\gamma}^{k'}(z_3 n) W^{k'k}(z_3 n, z_0 n) | B(P) \rangle,
\label{eq:baryon_matrix}
\end{aligned}
\end{equation}
where $| \Lambda(P, \lambda)\rangle$ denotes the $\Lambda$ baryon state with momentum $P$ and helicity $\lambda$. The indices $\alpha$, $\beta$, and $\gamma$ refer to Dirac components, while $f$, $g$, and $h$ represent the quark flavor fields. Two light-cone vectors are defined as $n^\mu = \tfrac{1}{\sqrt{2}}(1,0,0,-1)$ and $\bar n^\mu = \tfrac{1}{\sqrt{2}}(1,0,0,1)$, with the baryon momentum taken along the $\bar n$ direction, $P^\mu = P^+ \bar n^\mu = (P^z, 0, 0, P^z)$. The light-like Wilson lines \( W_{ij} \) connect the nonlocal quark fields to a common reference point \( z_0 \), ensuring gauge invariance of the operator. For simplicity, we omit the explicit Wilson lines in what follows.


The most general decomposition of Eq.~\eqref{eq:baryon_matrix} involves 24 invariant functions~\cite{Braun:2000kw}, but only three of them correspond to leading twist (twist~3):
\begin{equation}
\begin{split}
H(z_{1},z_{2},z_{3})_{\alpha\beta\gamma}
= \frac{1}{4} f_{V} \Bigl[ (\slashed{P} C)_{\alpha\beta} (\gamma_{5}u_{\Lambda})_{\gamma}
\Phi_V(z_{i} P\!\cdot\! n)& \\
\quad + (\slashed{P}\gamma_{5} C)_{\alpha\beta} (u_{\Lambda})_{\gamma}
\Phi_A(z_{i} P\!\cdot\! n) \Bigr] &\\
\quad + \frac{1}{4} f_{T} \Bigl[(i\sigma_{\mu\nu} P^{\nu} C)_{\alpha\beta}
(\gamma^{\mu}\gamma_{5} u_{\Lambda})_{\gamma}
\Phi_T(z_{i} P\!\cdot\! n)\Bigr]&.
\end{split}
\end{equation}

The individual leading-twist LCDAs can be projected by inserting appropriate Dirac matrices~$\Gamma$ in the nonlocal operator,
\begin{equation}
\begin{aligned}
\Phi_{V/A/T}(z_{1},&z_{2},\mu)\, P^{+} f_{A/T} u_{\Lambda}(P)\\
&= \,
\langle 0 | u^{ \mathrm{T}}(z_{1}n)\, \Gamma\,
d^(z_{2}n)\, s(0)
| \Lambda(P)\rangle,
\end{aligned}
\end{equation}
where $\mu$ denotes the renormalization scale.  
In this work, we primarily focus on the axial amplitude $A(z_i P\!\cdot\! n)$, corresponding to $\Gamma = C\, n\!\!\!/\gamma_5$.  
For simplicity, we set $z_3 = 0$.  
The momentum-space LCDA is then defined as
\begin{equation}
\begin{aligned}
\phi_{A}(x_{1},x_{2},\mu)
= \int &\frac{P^{+} dz_{1}}{2\pi} \int \frac{P^{+} dz_{2}}{2\pi}\,\\
&e^{i(x_{1} z_{1} + x_{2} z_{2}) P^{+}}\,
\Phi_{A}(z_{1},z_{2},\mu).
\end{aligned}
\end{equation}

The LCDAs can be extracted from the corresponding quasi-DAs via factorization. The quasi-DA for $A$-wave is defined by
\begin{equation}
\begin{split}
\tilde{\phi}_{A}(x_1,x_2,\mu)
= \int &\frac{P^z \, dz_1}{2\pi} \int \frac{P^z \, dz_2}{2\pi}\,\\
&e^{-i(x_1 z_1 + x_2 z_2) P^z}\,
\tilde{\Phi}_{A}(z_1, z_2, P^z, \mu),
\end{split}
\end{equation}
where $\tilde{\Phi}_{A}$ is the equal-time correlator,
\begin{equation}
\begin{aligned}
\tilde{\Phi}_A(z_1, z_2, P^z, \mu)\,& f_A P^{\nu} u_\Lambda(P)
=\\ \langle 0 |
&u^{T}(z_1 n_z)\,
(C \gamma_5 \gamma^\nu)\,
d(z_2 n_z)\,
s(0)
| \Lambda(P) \rangle,
\label{Awave}
\end{aligned}
\end{equation}
with $n_z^\mu = (0, 0, 0, -1)$ the unit vector along the spatial $z$ direction.  

The factorization relation between the quasi-DAs and LCDAs reads
\begin{equation}
\begin{aligned}
{\phi}_A(x_1, x_2, P^z, \mu)
&= \int_0^1 dy_1 \int_0^{1 - y_1} dy_2\,\\
&\qquad C(x_1, x_2, y_1, y_2, P^z, \mu)\,
\tilde\phi_A(y_1, y_2, \mu) \\
+ \mathcal{O}\!&\left(
\frac{\Lambda_{\text{QCD}}^2}{(x_1 P^z)^2},
\frac{\Lambda_{\text{QCD}}^2}{(x_2 P^z)^2},
\frac{\Lambda_{\text{QCD}}^2}{[(1 - x_1 - x_2) P^z]^2}
\right),
\label{lamet_fact}
\end{aligned}
\end{equation}
where the power corrections are suppressed in the large-momentum limit.  
The perturbative matching kernel can be expanded in $\alpha_s$ as
\begin{equation}
\begin{aligned}
C(x_1, x_2, y_1, y_2,& P^z, \mu)
= \delta(x_1 - y_1) \delta(x_2 - y_2) \\
& + \frac{\alpha_s C_F}{2\pi}\,
c^{(1)}(x_1, x_2, y_1, y_2, P^z, \mu)
+ \mathcal{O}(\alpha_s^2),
\end{aligned}
\end{equation}
where the explicit one-loop expression for $c^{(1)}$ can be found in Refs.~\cite{Deng:2023csv,Han:2023xbl,Han:2024ucv}.

Extracting baryon quasi-DAs on the lattice, however, remains highly challenging. Unlike the mesonic case, the baryon distribution amplitude depends on two independent momentum fractions, making it effectively a two-dimensional quantity. This feature significantly amplifies the signal-to-noise problem. In addition, the renormalization of such multi-variable nonlocal operators is considerably more involved. 

In our previous work, we developed a two-dimensional hybrid renormalization scheme to handle the associated mixing and divergence structure, with a detailed description provided in Refs.~\cite{Han:2023xbl,Han:2024ucv}. Despite its theoretical consistency, the approach remains numerically demanding. These challenges motivate the introduction of the gradient-flow formalism in the next subsection, which provides a framework to regularize and renormalize baryon quasi-DAs more efficiently.

\subsection{Gradient Flow Formalism}

The gradient flow provides a systematic framework for suppressing UV fluctuations in both gauge and fermion fields~\cite{Narayanan:2006rf,Luscher:2010iy,Luscher:2011bx,Luscher:2009eq}. By evolving the fields along an auxiliary fifth dimension, the flow time \(t\), short-distance modes are smoothed while long-range physics is preserved. Owing to these properties, the gradient flow has become an essential tool in lattice QCD, widely applied to scale setting, renormalization, and improving signal quality in hadronic observables~\cite{Luscher:2013cpa,Luscher:2013vga,BMW:2012hcm,Sommer:2014mea,Suzuki:2013gza,Makino:2014taa,Harlander:2018zpi,FlavourLatticeAveragingGroupFLAG:2024oxs,Leino:2021vop,Mayer-Steudte:2022uih,Leino:2022kgj}.

The flow transforms the gauge field \(A_\mu(x)\) and fermion field \(\psi(x)\) into flowed fields \(B_\mu(t,x)\) and \(\chi(t,x)\), respectively, obtained by solving diffusion-like equations:
\begin{equation}
\begin{aligned}
\partial_t B_\mu &= D_\nu G_{\nu\mu} + \kappa D_\mu \partial_\nu B_\nu, \\
\partial_t \chi &= D_\mu D_\mu \chi - \kappa (\partial_\mu B_\mu)\chi, \\
\partial_t \bar{\chi} &= \bar{\chi}\overleftarrow{D}_\mu \overleftarrow{D}_\mu + \kappa \bar{\chi}\partial_\mu B_\mu,
\end{aligned}
\end{equation}
where \(D_\mu\) acts in the adjoint representation for gauge fields and in the fundamental representation for fermions. The parameter \(\kappa\) is a gauge-fixing term that does not affect physical observables. The flowed field-strength tensor is
\begin{equation}
G_{\mu\nu} = \partial_\mu B_\nu - \partial_\nu B_\mu + [B_\mu, B_\nu],
\end{equation}
with boundary conditions
\begin{equation} \begin{aligned} B_\mu(x; t=0) &= g A_\mu(x), \\ \chi(x; t=0) &= \psi(x), \\ \bar{\chi}(x; t=0) &= \bar{\psi}(x). \end{aligned} \end{equation}

\begin{@twocolumnfalse}
\begin{center}
\begin{figure*}[t]
    \centering
    \includegraphics[width=0.9\linewidth]{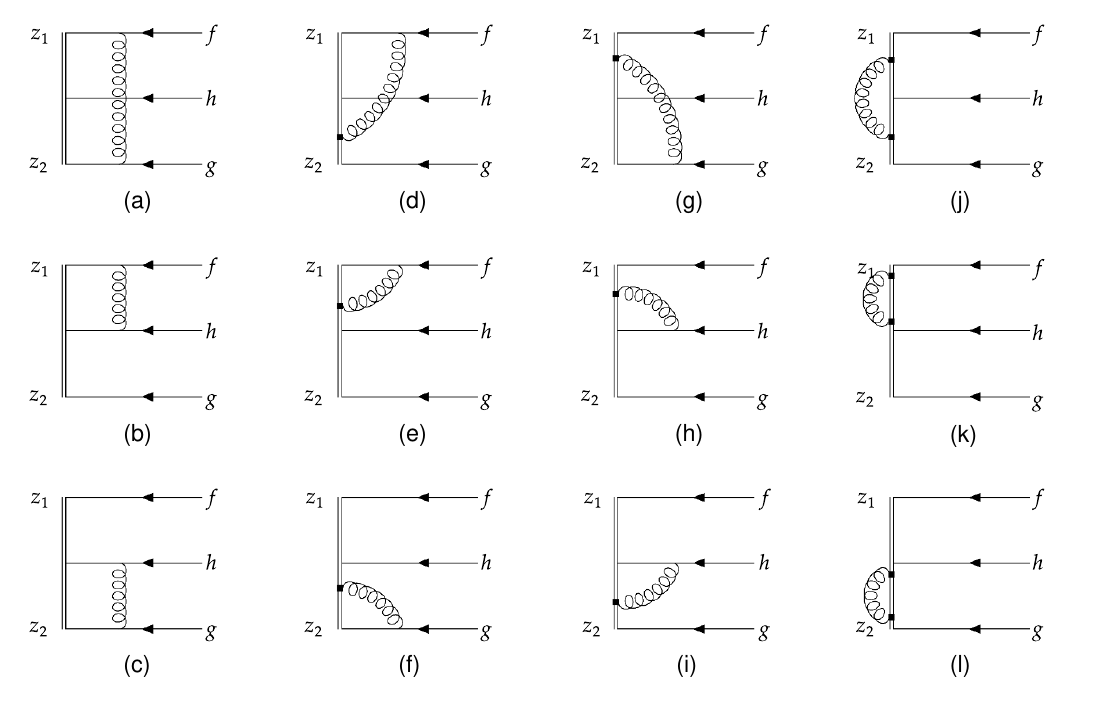}
\caption{One-loop contributions to the equal-time correlator 
$\tilde{\Phi}^{R}(z_1, z_2, P^z, t)$. 
The double line represents the Wilson line, and the filled squares mark the flowed gauge fields 
$B_\mu$ at flow time~$t$.}
    \label{quasi_diagrams}
\end{figure*}
\end{center}
\end{@twocolumnfalse}

As the flow time increases, UV fluctuations are exponentially damped and the fields become smooth over a characteristic length scale \(\sqrt{8t}\). Composite operators built from flowed fields, referred to as ``flowed operators'' are automatically UV finite at nonzero \(t\), eliminating the need for additional counterterms. This property makes the gradient flow particularly powerful for defining renormalized operators on the lattice.

Moreover, the flow introduces a physical scale \(\sqrt{8t}\) that can serve as a nonperturbative renormalization scale or as a matching scale between lattice and continuum schemes. Thus, the gradient flow simultaneously improves statistical precision and provides theoretical control over renormalization and scheme conversion.

Unlike smearing techniques such as stout smearing~\cite{Luscher:2009eq,Nagatsuka:2023jos}, which obscure perturbative structures, the gradient flow preserves the applicability of perturbation theory. The gradient flow formalism admits a well-defined expansion with modified Feynman rules~\cite{Lange:2021vqg}, maintaining compatibility with continuum calculations.

In the following sections, we apply the gradient-flow formalism to baryon quasi-distribution amplitudes. It serves as a replacement for self-renormalization, offering a gauge-invariant and UV-finite framework that enables a controlled conversion from the lattice scheme to the $\overline{\mathrm{MS}}$ scheme and a systematic elimination of short-distance divergences.

\section{Factorization of Baryon Quasi-DAs with Gradient Flow}

The factorization of flowed quasi-observables into their $\overline{\mathrm{MS}}$ counterparts has been firmly established in the study of quasi-PDFs~\cite{Brambilla:2023vwm, Monahan:2017hpu,Monahan:2016bvm}. In these analyses, the gradient flow serves as a gauge-invariant framework that enables a well-defined matching of lattice observables to the $\overline{\mathrm{MS}}$ scheme. An important outcome of these studies is that, in the small flow-time limit $t \to 0$, the matching kernel becomes free from any explicit dependence on the spatial interval.

In this section, we establish the factorization relation that connects the flowed baryon quasi-DAs to their counterparts renormalized in the $\overline{\mathrm{MS}}$ scheme. The factorization formula reads
\begin{equation}
\label{flow_match}
\tilde \Phi^{R}(z_1,z_2,P^z,t) =e^{\delta m\tilde z} {\cal C}_{q}(t, z_1,z_2, \mu) \, \tilde\Phi^{\overline{\mathrm{MS}}}(z_1,z_2,P^z,\mu),
\end{equation}
where $\tilde\Phi^{R}(z_1,z_2,P^z,t)$ stands for the flowed quasi-DA. The factor \( e^{\delta m \tilde{z}} \) represents the linear divergence arising from the self-energy corrections of the Wilson line, where
\begin{equation}
\label{self_energy}
\delta m = -\frac{\alpha_s}{4\pi}C_F\frac{\sqrt{2\pi}}{\sqrt{t}} + \mathcal{O}(\alpha_s^2),
\end{equation}
and
\begin{equation}
\label{z}
\tilde{z} = \begin{cases}
|z_1 - z_2|, & \text{if } z_1 z_2 < 0, \\
\max(|z_1|, |z_2|), & \text{if } z_1 z_2 \geq 0.
\end{cases}
\end{equation}
This matching relation factorizes the small \(t\) effects into the perturbative coefficient \(\mathcal{C}_q(t,z_1,z_2,\mu)\), thereby separating the flow-induced UV structure from the long-distance dynamics contained in \(\tilde{\Phi}^{\overline{\mathrm{MS}}}(z_1,z_2,P^z,\mu)\).

In the $\overline{\mathrm{MS}}$ scheme, the Wilson-line self-energy contains the UV renormalon~\cite{Beneke:1998ui,Beneke:1994sw}, which manifests as a factorial growth of the perturbative coefficients at large orders. In contrast, within the gradient-flow formalism, a finite flow time \( t \) effectively removes the UV renormalon associated with this linear divergence~\cite{Zhang:2025jmq}.

Ref.~\cite{Brambilla:2023vwm} suggests that, in the small flow-time limit, the matching kernel can be derived from the matching of local composite operators using the auxiliary-field method, provided that the matrix elements satisfy the condition \( z \gg \sqrt{8t} \). However, for baryon LCDAs, which involve three spatial separations, this hierarchy is not guaranteed throughout the entire configuration space. Consequently, there exist regions where the required separation between physical scales is violated, potentially introducing uncontrolled systematic effects.

To improve the precision and reliability of the matching, we compute the full one-loop matching kernel beyond the small flow-time approximation.

In the following, we focus on the calculation of the $A$-wave of the flowed quasi-DA for the $\Lambda$ baryon.
The one-loop diagrams contributing to $\tilde\Phi^{R}(z_1,z_2,P^z,t)$ are shown in Fig.~\ref{quasi_diagrams}. Corrections to the external legs are omitted in the calculation since they cancel in the ratio scheme, which will be discussed in Sec.~IV.

\subsection{One-loop corrections for the flowed quasi-DA}
 Having established the factorization relation for baryon quasi-DAs with gradient flow, we now proceed to the explicit one-loop calculation of the matching kernel and the Wilson-line linear divergence. The calculation is performed using dimensional regularization in $d = 4 - 2\epsilon$ dimensions, and the $\overline{\mathrm{MS}}$ scheme is adopted for renormalization. The gauge parameters are set to $\kappa=1$ and $\xi=1$ for simplicity.
 
 To determine the matching kernel, one needs the one-loop corrections to the quasi-DA computed both in the gradient-flow scheme and in the $\overline{\mathrm{MS}}$ scheme. The difference between these two results defines the perturbative matching coefficient ${\cal C}_q(t,z_1,z_2,\mu)$ in Eq.~\eqref{flow_match}.
Because the factorization removes only the short-distance dependence on the small flow time \(t\), the external quark momenta, which belong to the infrared dynamics, do not enter the perturbative matching kernel. Consequently, we evaluate ${\cal C}_q(t,z_1,z_2,\mu)$ with vanishing external momenta to simplify the calculation without affecting the result.
Therefore, we first present the individual one-loop results in the gradient-flow and $\overline{\mathrm{MS}}$ scheme.

We begin with the one-loop flowed gluon-exchange diagrams between external quarks, hereafter referred to as the ``box diagrams,'' shown in Fig.~\ref{quasi_diagrams}(a)--(c). The corresponding contributions to the flowed matrix elements are
\begin{equation}
\begin{aligned}
\tilde\Phi^R_a(z_1,z_2,P^z,t)=&\frac{ C_F\alpha_s}{8\pi}\left[-\frac{1}{\epsilon _{\text{IR}}}-\log \left(\frac{1}{4}\mu ^2z_{12}^2e^{2\gamma_E-3}\right) \right.\\
+\operatorname{Ei}\left(-\bar{z}_{12}^2\right)&\left.+\frac{3}{\bar{z}^4}(1-e^{-\bar{z}_{12}^2})-\frac{1}{\bar{z}^2}(4-e^{-\bar{z}_{12}^2})\right],  \\
\tilde\Phi^R_b(z_1,z_2,P^z,t)=&\frac{C_F\alpha_s }{16 \pi}\left[-\frac{1}{\epsilon _{\text{IR}}}- \log \left(\frac{1}{4}\mu^2  z_1^2e^{2\gamma_E-1}\right)\right.\\
&\left.+\operatorname{Ei}\left(-\bar{z}_1^2\right)-\frac{2 }{\bar{z}_1^2}\left(1-e^{-\bar{z}_1^2}\right)\right],\\
\tilde\Phi^R_c(z_1,z_2,P^z,t)=&\frac{C_F\alpha_s }{16 \pi}\left[-\frac{1}{\epsilon _{\text{IR}}}- \log \left(\frac{1}{4}\mu^2  z_2^2e^{2\gamma_E-1}\right)\right.\\
&\left.+\operatorname{Ei}\left(-\bar{z}_2^2\right)-\frac{2 }{\bar{z}_2^2}\left(1-e^{-\bar{z}_2^2}\right)\right],
\end{aligned}
\end{equation}
while the corresponding expressions in the $\overline{\mathrm{MS}}$ scheme are
\begin{equation}
\begin{aligned}
\tilde\Phi^{\overline{\mathrm{MS}}}_a(z_1,z_2,P^z,\mu)=&-\frac{C_F\alpha_s}{8\pi \epsilon_{\text{IR}}}-\frac{C_F \alpha _s  \log \left(\frac{1}{4}\mu ^2 z_{12}^2e^{2\gamma_E-3}\right)}{8 \pi  }, \\
\tilde\Phi^{\overline{\mathrm{MS}}}_b(z_1,z_2,P^z,\mu)=&-\frac{C_F\alpha_s}{16\pi \epsilon_{\text{IR}}}-\frac{C_F \alpha _s \log\left(\frac{1}{4}\mu^2z_1^2e^{2\gamma_E-1}\right)}{16 \pi }, \\
\tilde\Phi^{\overline{\mathrm{MS}}}_c(z_1,z_2,P^z,\mu)=&-\frac{C_F\alpha_s}{16\pi \epsilon_{\text{IR}}}-\frac{C_F \alpha_s \log\left(\frac{1}{4}\mu^2z_2^2e^{2\gamma_E-1}\right)}{16 \pi }.
\end{aligned}
\end{equation}
Here, \( z_{12} \equiv z_1 - z_2 \) and \( \bar{z}_i \equiv z_i/\sqrt{8t} \). In the small flow-time limit \( t \to 0 \), the flowed matrix elements approach their $\overline{\mathrm{MS}}$ counterparts. This agrees with expectations, since the matching kernel in this limit coincides with that of three local operators, as described in the auxiliary-field formalism~\cite{Brambilla:2023vwm}.

Fig.~\ref{quasi_diagrams}(d) and Fig.~\ref{quasi_diagrams}(g) yield no contribution to the matching kernel in the small flow-time limit for the same reason. Their flowed results are

\begin{equation}
\begin{aligned}
\tilde\Phi^R_d(z_1,z_2,&P^z,t)=\frac{C_F \alpha _s }{8 \pi }\bigg[\log
\left(\bar{z}_{12}^2\right)-\log \left(\bar{z}_1^2\right)\\
&-\operatorname{Ei}\left(-\bar{z}_{12}^2\right)+\operatorname{Ei}\left(-\bar{z}_1^2\right)\\
&+\frac{1}{\bar{z}_{12}^2}\left( 1-e^{-\bar{z}_{12}^2}\right)-\frac{ 1}{\bar{z}_1^2}\left(1-e^{-\bar{z}_1^2}\right)\bigg]\\
\tilde\Phi^R_g(z_1,z_2,&P^z,t)=\frac{C_F \alpha _s }{8 \pi }\bigg[\log
\left(\bar{z}_{12}^2\right)-\log \left(\bar{z}_2^2\right)\\
&-\operatorname{Ei}\left(-\bar{z}_{12}^2\right)+\operatorname{Ei}\left(-\bar{z}_2^2\right)\\
&+\frac{1}{\bar{z}_{12}^2}\left( 1-e^{-\bar{z}_{12}^2}\right)-\frac{ 1}{\bar{z}_2^2}\left(1-e^{-\bar{z}_2^2}\right)\bigg],
\end{aligned}
\end{equation}
with the corresponding $\overline{\mathrm{MS}} $ results
\begin{equation}
\begin{aligned}
\tilde\Phi^{\overline{\mathrm{MS}}}_d(z_1,z_2,&P^z,\mu)=\\
&\frac{C_F \alpha _s }{8 \pi }\bigg[\log \left(z_{12}^2\mu^2e^{2\gamma_E}\right)-\log \left(z_1^2\mu^2e^{2\gamma_E}\right)\bigg],\\
\tilde\Phi^{\overline{\mathrm{MS}}}_g(z_1,z_2,&P^z,\mu)=\\
&\frac{C_F \alpha _s }{8 \pi }\bigg[\log \left(z_{12}^2\mu^2e^{2\gamma_E}\right)-\log \left(z_2^2\mu^2e^{2\gamma_E}\right)\bigg].
\end{aligned}
\end{equation}
In the remainder of this subsection, we present the results for the remaining diagrams. Unlike the box and certain vertex diagrams, these contributions do not reduce to their corresponding expressions in the $\overline{\mathrm{MS}}$ scheme in the small flow-time limit. Consequently, a nontrivial matching coefficient is required to relate the flowed quasi-DAs to their $\overline{\mathrm{MS}}$ counterparts.
The results for Fig.~\ref{quasi_diagrams}(e) and Fig.~\ref{quasi_diagrams}(f) are
\begin{equation}
\begin{aligned}
\tilde\Phi^R_e(z_1,z_2,P^z,t)=&\frac{C_F\alpha_s}{4\pi}\bigg[ \log \left(\bar{z}_1^2e^{\gamma_E-1}\right)\\
&-Ei\left(-\bar{z}_1^2\right)+\frac{1}{\bar{z}_1^2}\left(1-e^{-\bar{z}_1^2}\right)\bigg],\\
\tilde\Phi^R_f(z_1,z_2,P^z,t)=&\frac{C_F\alpha_s}{4\pi}\big[ \log \left(\bar{z}_2^2e^{\gamma_E-1}\right)\\
&-Ei\left(-\bar{z}_2^2\right)+\frac{1}{\bar{z}_2^2}\left(1-e^{-\bar{z}_2^2}\right)\bigg],
\end{aligned}
\end{equation}
with the corresponding $\overline{\mathrm{MS}} $ results
\begin{equation}
\begin{aligned}
\tilde\Phi^{\overline{\mathrm{MS}}}_e(z_1,z_2,P^z,\mu)=&\frac{C_F\alpha_s}{4\pi}\log\left(\frac{1}{4}  \mu ^2 z_1^2e^{2 \gamma_E }\right),\\
\tilde\Phi^{\overline{\mathrm{MS}}}_f(z_1,z_2,P^z,\mu)=&\frac{C_F\alpha_s}{4\pi}\log\left(\frac{1}{4}\mu ^2 z_2^2 e^{2 \gamma_E }\right).
\end{aligned}
\end{equation}

The results for Fig.~\ref{quasi_diagrams}(h) and Fig.~\ref{quasi_diagrams}(i) are
\begin{equation}
\begin{aligned}
\tilde\Phi^R_h(z_1,z_2,P^z,t)=&\frac{C_F \alpha _s }{8\pi}\bigg[ \log \left(\bar{z}_1^2e^{\gamma_E-1}\right)\\
&-\operatorname{Ei}(-\bar{z}_1^2)+\frac{1}{ \bar{z}_1^2}(1- e^{-\bar{z}_1^2})\bigg]\\
\tilde\Phi^R_i(z_1,z_2,P^z,t)=&\frac{C_F \alpha _s }{8\pi}\bigg[ \log \left(\bar{z}_2^2e^{\gamma_E-1}\right)\\
&-\operatorname{Ei}(-\bar{z}_2^2)+\frac{1}{ \bar{z}_2^2}(1- e^{-\bar{z}_2^2})\bigg],
\end{aligned}
\end{equation}
with the corresponding $\overline{\mathrm{MS}} $ results
\begin{equation}
\begin{aligned}
\tilde\Phi^{\overline{\mathrm{MS}}}_h(z_1,z_2,P^z,\mu)=&\frac{C_F\alpha_s}{8\pi}\log\left(\frac{1}{4}\mu ^2 z_1^2e^{2 \gamma_E } \right),\\
\tilde\Phi^{\overline{\mathrm{MS}}}_i(z_1,z_2,P^z,\mu)=&\frac{C_F\alpha_s}{8\pi}\log\left(\frac{1}{4} \mu ^2 z_2^2e^{2 \gamma_E } \right).
\end{aligned}
\end{equation}

Finally, we present the Wilson line corrections Fig.~\ref{quasi_diagrams}(j), Fig.~\ref{quasi_diagrams}(k), Fig.~\ref{quasi_diagrams}(l)
\begin{equation}
\begin{aligned}
\tilde\Phi^R_j(z_1,z_2,P^z,t)=\frac{C_F \alpha_s }{4 \pi }\bigg[2
   \left(e^{-\bar{z}_1^2}+e^{-\bar{z}_2^2}-e^{-\bar{z}_{12}^2}\right)&\\
+2 \sqrt{\pi } \bigg(\bar{z}_1\text{erf}(\bar{z}_1)+\bar{z}_2\text{erf}(\bar{z}_2)-\bar{z}_{12}\text{erf}(\bar{z}_{12})\bigg)&\\
-\operatorname{Ei}\left
(-\bar{z}_{12}^2\right)+\operatorname{Ei}\left(-\bar{z}_1^2\right)+\operatorname{Ei}\left(-\bar{z}_2^2\right)&\\
-\log \left(\bar{z}_1^2\right)-\log \left(\bar{z}_2^2\right)+\log
\left(\bar{z}_{12}^2\right)-\gamma_E -2\bigg]&,\\
\tilde\Phi^R_k(z_1,z_2,P^z,t)=\frac{C_F\alpha_s}{2\pi }\bigg[-2 e^{-\bar{z}_1^2}-2 \sqrt{\pi } \bar{z}_1 \text{erf}(\bar{z}_1)&\\
-\operatorname{Ei}\left(-\bar{z}_1^2\right)+2 \log (\bar{z}_1)+\gamma_E +2\bigg]&,\\
\tilde\Phi^R_l(z_1,z_2,P^z,t)=\frac{C_F\alpha_s}{2\pi }\bigg[-2 e^{-\bar{z}_2^2}-2 \sqrt{\pi } \bar{z}_2 \text{erf}(\bar{z}_2)&\\
-\operatorname{Ei}\left(-\bar{z}_2^2\right)+2 \log (\bar{z}_2)+\gamma_E +2\bigg]&,
\end{aligned}
\end{equation}
with the corresponding $\overline{\mathrm{MS}} $ results
\begin{equation}
\begin{aligned}
\tilde\Phi^{\overline{\mathrm{MS}}}_j(z_1,&z_2,P^z,\mu)=\frac{C_F \alpha _s }{4 \pi }\bigg[\log \left(\frac{1}{4}\mu^2z_{12}^2e^{2\gamma_E}\right)\\
&-\log \left(\frac{1}{4}\mu^2z_1^2e^{2\gamma_E}\right)-\log \left(\frac{1}{4}\mu^2z_2^2e^{2\gamma_E}\right)-2\bigg],\\
\tilde\Phi^{\overline{\mathrm{MS}}}_k(z_1,&z_2,P^z,\mu)=\frac{ C_F\alpha_s}{2\pi} \log\left( \frac{1}{4}z_1^2e^{2\gamma_E+2} \right),\\
\tilde\Phi^{\overline{\mathrm{MS}}}_l(z_1,&z_2,P^z,\mu)=\frac{ C_F\alpha_s}{2\pi} \log\left( \frac{1}{4}z_2^2e^{2\gamma_E+2} \right).
\end{aligned}
\end{equation}

The linear divergence originating from the Wilson-line self-energy becomes manifest in the gradient-flow formalism, where it is associated with the error function,
\begin{equation}
\operatorname{erf}(\bar{z}) = \frac{2}{\sqrt{\pi}} \int_0^{\bar{z}} e^{-t^2} dt,
\end{equation}
which arises from the Gaussian smearing induced by the flow. In the limit of vanishing flow time, the relation
\begin{equation}
\lim_{t \to 0} \bar{z} \, \operatorname{erf}(\bar{z}) = |\bar{z}|
\end{equation}
holds, reproducing the usual linear divergence structure. These terms correspond to the mass counterterm \( \delta m \) introduced in Eq.~\eqref{flow_match}. The total linear divergence from the self-energy diagrams is given by
\begin{equation}
\delta m \tilde z = -\frac{C_F \alpha_s}{2\sqrt{\pi}} \left( \bar{z}_1 \operatorname{erf}(\bar{z}_1) + \bar{z}_2 \operatorname{erf}(\bar{z}_2) + \bar{z}_{12} \operatorname{erf}(\bar{z}_{12}) \right).
\end{equation}
In the small flow-time limit, this expression simplifies to
\begin{equation}
\lim_{t \to 0} \delta m \tilde z = -\frac{C_F \alpha_s}{4\pi} \frac{\sqrt{2\pi}}{\sqrt{t}} \cdot \frac{|z_1| + |z_2| + |z_{12}|}{2}.
\end{equation}
The combination \( \frac{|z_1| + |z_2| + |z_{12}|}{2} \) corresponds to the effective length \( \tilde{z} \) defined in Eq.~\eqref{z}, confirming the consistency of the flowed self-energy correction with Eq.~\eqref{self_energy}.

In addition, although the external quark self-energy cancels in our final procedure defined in Sec.~IV, it is instructive to present it here for completeness and for comparison with the result derived from the auxiliary field method. The flowed quark self-energy reads
\begin{equation}
\label{self_energy}
\mathcal{C}_\chi(\mu,t) = 1 + \frac{3}{2} \frac{\alpha_s}{4\pi} C_F 
\left[ \log\left(2\mu^2 t e^{\gamma_E}\right) - \log(432) \right].
\end{equation}

\subsection{Full Matching Kernel}

\begin{figure*}
\centering
\begin{minipage}[t]{0.47\textwidth}
\centering
\includegraphics[width=8cm]{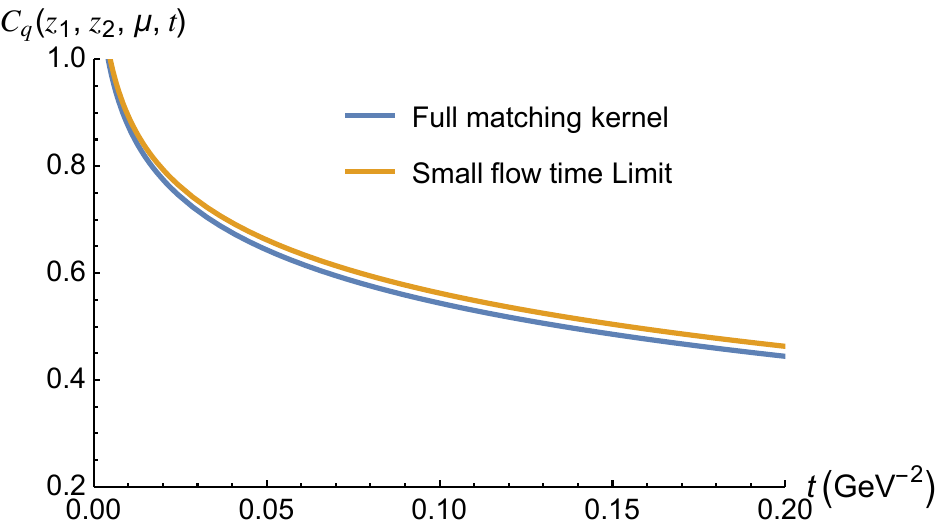}
\centerline{(a)}
\end{minipage}
\begin{minipage}[t]{0.47\textwidth}
\centering
\includegraphics[width=7cm]{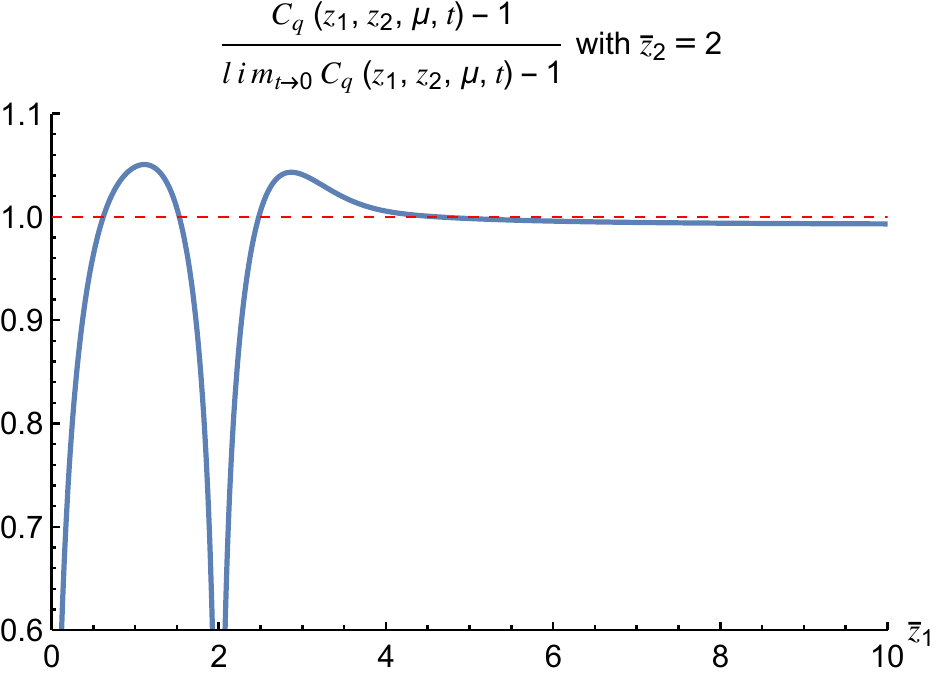}
\centerline{(b)}
\end{minipage}
\caption{(a) Variation of the matching kernel with the flow time \(t\) for fixed parameters \( \bar{z}_1 = 1 \), \( \bar{z}_2 = 2 \), and \( \mu = 2~\mathrm{GeV} \). The blue curve shows the full matching kernel, while the red curve indicates its small flow-time limit.
(b) \( \dfrac{C_q(z_1, z_2, \mu, t) - 1}{\lim_{t \to 0} C_q(z_1, z_2, \mu, t) - 1} \), showing the ratio of the one-loop correction of the full matching kernel to its small flow-time limit, with \( \bar{z}_2 = 2 \) and \( t = 0.03~\mathrm{fm}^2 \). The renormalization scale \( \mu \) is chosen such that \( \log\bigl(2\mathrm{e}^{2\gamma_E} \mu^2 t\bigr) = 0 \).}
\label{compare}
\end{figure*}

Having obtained the one-loop corrections to the flowed quasi-DA and its $\overline{\mathrm{MS}}$ counterpart, we now present the resulting matching kernel that connects the two schemes. The full matching kernel at finite flow time reads:

\vspace{1em}

\begin{widetext}
\begin{equation}
\label{full_kernel}
\begin{aligned}
\mathcal{C}_q(z_1,z_2,t,\mu)
=&1-\frac{C_F \alpha _s }{16 \pi }\bigg[18\log\left(2\mu^2te^{\gamma_E}\right)+6\log(432)+12+7 \operatorname{Ei}\left(-\bar z_1^2\right)+7 \operatorname{Ei}\left(-\bar z_2^2\right)+6
   \operatorname{Ei}\left(-\bar z_{12}^2\right)\\
   &+8
   \left(e^{-\bar z_1^2}+e^{-\bar z_2^2}+e^{-\bar z_{12}^2}\right)
   +\frac{2
   \left(e^{-\bar z_1^2}-1\right)}{\bar z_1^2}+\frac{2
   \left(e^{-\bar z_2^2}-1\right)}{\bar z_2^2}+\frac{6\left( e^{-\bar z_{12}^2}-1\right)}{\bar z_{12}^4}+\frac{2\left( e^{-\bar z_{12}^2}+2\right)}{\bar z_{12}^2}\bigg]+\mathcal{O}(\alpha_s^2),
\end{aligned}
\end{equation}
\end{widetext}

and in the small flow-time limit it reduces to
\begin{equation}
\begin{aligned}
\label{eq:kernellimit}
\lim_{t\to 0}&\,\mathcal{C}_q(z_1,z_2,t,\mu)
=\\
&1 - \frac{3 C_F\alpha_s}{8\pi} \bigg[3 \log\left(2\mu^2 t e^{\gamma_E}\right) + 2+\log(432) \bigg] + \mathcal{O}(\alpha_s^2).
\end{aligned}
\end{equation}

It has been shown in Ref.~\cite{Brambilla:2023vwm} that, by introducing auxiliary fields, the matching of a nonlocal operator in the small flow-time limit can be reduced to that of local operators. In particular, the equal-time operator in Eq.~(\ref{Awave}) can be rewritten as
\begin{equation}
\bar{h}_v u_\alpha(z_1 n_z)\, (C n\!\!\!/\gamma_5)_{\alpha\beta}\, \bar{h}_v d_\beta(z_2 n_z)\, \bar{h}_v s_\gamma(z_3 n_z),
\end{equation}
where \( h_v(x) \) is the auxiliary quark field that effectively replaces the Wilson line. The matching kernel then depends only on the local bilinear combinations \( \bar{h}_v \psi(z_i n_z) \) and is independent of the spinor structure or spatial separations. The coefficient for a single \( \bar{h}_v \psi(z_i n_z) \) takes the form
\begin{equation}
\begin{aligned}
\lim_{t \to 0} &\,\mathcal{C}_{\bar{h}_v \psi}(z_1, z_2, t,\mu)
= \\
&1 - \frac{C_F\alpha_s}{8\pi} \bigg[3 \log\left(2\mu^2 t e^{\gamma_E}\right) + 2+\log(432) \bigg] + \mathcal{O}(\alpha_s^2),
\end{aligned}
\end{equation}
which is exactly one-third of Eq.~(\ref{eq:kernellimit}). 

However, for baryon LCDAs, which depend on two independent spatial separations, the hierarchy \( |z_1|, |z_2|, |z_1 - z_2| \gg \sqrt{8t} \) is not always satisfied. Consequently, deviations arise in regions where this separation of scales breaks down. Fig.~\ref{compare} compares the full matching kernel with its small-flow-time limit. In Fig.~\ref{compare}(a), we fix $\bar{z}_1=1$ and $\bar{z}_2=2$ and vary \( t \) from \( 0 \) to \( 0.2~\mathrm{fm}^2 \); in Fig.~\ref{compare}(b), we show the ratio \( \mathcal{C}_q(z_1,z_2,\mu,t)/\lim_{t\to 0}\mathcal{C}_q(z_1,z_2,\mu,t) \) as a function of \( \bar{z}_1 \) for \( t=0.03~\mathrm{fm}^2 \). The renormalization scale \( \mu \) is chosen such that \( \log(2e^{2\gamma_E}\mu^2t)=0 \).

One can observe two peaks where the ratio significantly deviates from unity, corresponding to the cases \( z_1 = z_2 \) and \( z_2 = 0 \). However, these deviations can be removed in the ratio scheme, provided that the dependence on the coordinates \( z_i \) factorizes to all orders. In particular, if such factorization holds, the flowed quasi-DAs in coordinate space, in the limit of small spatial separation, can be written in the form \( f_1(z_1)\, f_2(z_2)\, f_3(z_1 - z_2) \).  The ratio scheme then cancels the functions \( f_i(z_i) \) for small \( z_i \) by forming ratios with matrix elements that share the same functional dependence, effectively eliminating these contributions.

\begin{@twocolumnfalse}
\begin{center}
\begin{figure*}[t]
    \centering
    \includegraphics[width=0.8\linewidth]{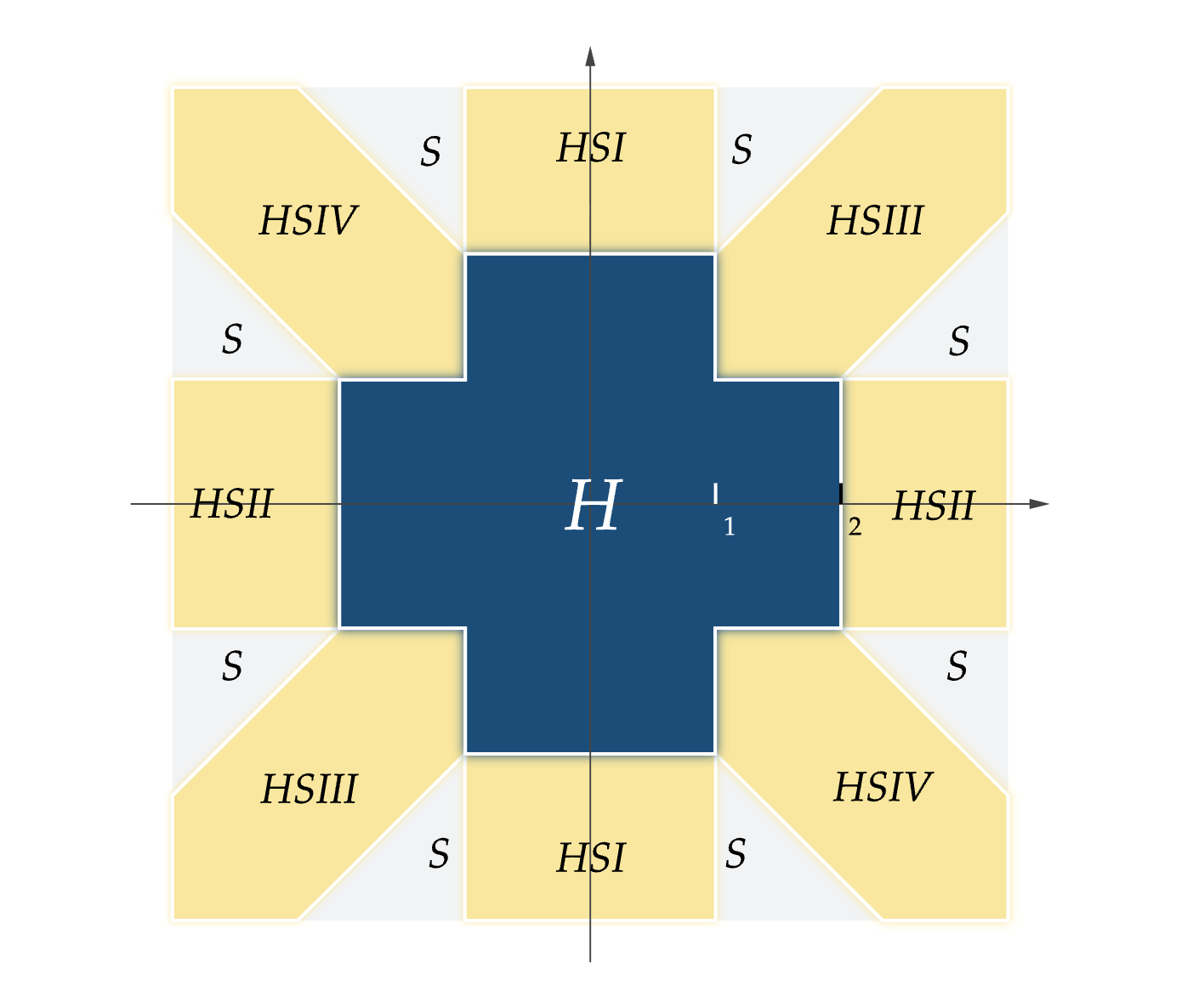}
    \caption{Schematic diagram of the renormalization procedure. First, both large-momentum and zero-momentum matrix elements are converted to the $\overline{\mathrm{MS}}$ scheme using the gradient-flow formalism. Then, ratios are formed using these $\overline{\mathrm{MS}}$-renormalized matrix elements: the large-momentum matrix element in the blue (hard) region is divided by the zero-momentum element in the same region; the large-momentum matrix element in the yellow (hard-soft) region is divided by the zero-momentum element on the blue-yellow boundary; and the large-momentum matrix element in the gray (soft) region is divided by the zero-momentum element at the gray-yellow intersection.}
    \label{fig:renorm}
\end{figure*}
\end{center}
\end{@twocolumnfalse}

\section{Hybrid Renormalization}

In Section~II, we introduced the factorization relation connecting Baryon quasi-DAs and LCDAs. We note that the matching in Eq.~(\ref{lamet_fact}) for quasi-DAs in the ${\overline{\mathrm{MS}}}$ scheme presents a subtle problem. This issue arises from the structure of the double plus function in the matching kernel, which contains integrals over the momentum fractions $x_1'$ and $x_2'$ originating from the virtual corrections to DAs~\cite{Deng:2023csv}. In the limits $x_1', x_2' \to \infty$, these integrals behave asymptotically as $\int dx'/x'$, leading to logarithmic UV divergences. Although the complete matching kernel should include both real and virtual contributions to cancel these divergences, previous derivations often relied solely on the real diagrams, assuming that the virtual terms could be reconstructed through integration. As a result, a residual UV divergence remains, rendering a problematic definition of the double plus function.

In Ref.~\cite{Izubuchi:2018srq}, a properly defined plus function for quasi-PDFs was introduced by performing subtractions at infinite momentum. For quasi-DAs, Ref.~\cite{Deng:2023csv} proposed using the RI/MOM renormalization scheme to subtract these asymptotic contributions, making the plus function well defined. However, as shown in Ref.~\cite{Zhang:2024omt}, it requires precise gauge fixing to properly control systematic uncertainties.

To eliminate large logarithms and convert lattice regularization to a continuum scheme, the hybrid renormalization scheme was introduced. It combines the ratio scheme and the self-renormalization scheme. The ratio scheme removes large logarithmic terms such as $\ln z_1^2$, $\ln z_2^2$, and $\ln (z_1 - z_2)^2$ that appear as $z_i \to 0$~\cite{Orginos:2017kos,Radyushkin:2017cyf,Radyushkin:2017lvu}. In this approach, bare matrix elements are divided by corresponding zero-momentum matrix elements according to the spatial region $(z_1, z_2)$. The division of regions is illustrated in Fig.~\ref{fig:renorm}, and a detailed application of this scheme to the baryon case can be found in Ref.~\cite{Han:2023xbl}.

The self-renormalization factor, on the other hand, converts lattice-regularized matrix elements to the $\overline{\mathrm{MS}}$ scheme~\cite{LatticePartonLPC:2021gpi}. It is given by
\begin{equation} \begin{aligned} \label{eq:ZR} Z_{R}&(z_1,z_2,a,\mu) = \exp\Bigg[ \left(\frac{k}{a \ln[a \Lambda_{\rm QCD}]} - m_{0}\right) \tilde{z} \\ &+ \frac{\gamma_0}{b_0} \ln \left(\frac{\ln [1 /(a \Lambda_{\rm QCD})]}{\ln [\mu / \Lambda_{\overline{\rm MS}}]}\right) + \ln \left(1+\frac{d}{\ln (a \Lambda_{\rm QCD})}\right) \\ &+ f(z_1,z_2)\, a \Bigg]\, , \end{aligned} \end{equation}
where the parameters $k$, $m_0$, $\Lambda_{\mathrm{QCD}}$, $f(z_1,z_2)$, and $d$ are determined from zero-momentum matrix elements.

Using this renormalization factor, the lattice matrix element can be converted to the $\overline{\mathrm{MS}}$ scheme over the full kinematic range as
\begin{equation}
\label{eq:latinMSbar}
\tilde{\Phi}^{\overline{\mathrm{MS}}}\left(z_1, z_2, P^z, \mu\right) 
= \frac{\tilde{\Phi}\left(z_1, z_2, P^z, a\right)}{Z_{R}(z_1,z_2,a,\mu)}.
\end{equation}
Although $Z_{R}$ is extracted from zero-momentum matrix elements, it can be applied to large-momentum matrix elements since the renormalization factor is independent of the external states.

Although the conventional hybrid scheme has been successfully applied in previous lattice studies~\cite{LPC:2025jvd}, it still suffers from several limitations in practice. 
First, due to the two-dimensional nature of baryon quasi-DAs, the extraction of matrix elements from lattice simulations suffers from a poor signal-to-noise ratio, especially at large separations in the two-dimensional coordinate space. 
Second, the self-renormalization procedure requires fitting the parameters in Eq.~(\ref{eq:ZR}), which introduces additional systematic uncertainties and dependence on the chosen fitting range.

To overcome these issues, we propose an improved hybrid renormalization scheme in which gradient flow matching replaces self renormalization. This substitution improves the signal-to-noise ratio of nonlocal correlators and enables a direct, theoretically controlled matching to the $\overline{\mathrm{MS}}$ scheme. In the next subsection, we formalize the gradient-flow based hybrid renormalization and detail its implementation in lattice simulations.

\subsection{Hybrid renormalization with gradient flow}

Building on the one-loop factorization with gradient flow, we now specify the hybrid renormalization used in our analysis. In this scheme, the scheme conversion is performed directly with the flowed matching relation, avoiding the fit-dependent self-renormalization. Using Eq.~(\ref{flow_match}) together with Eq.~(\ref{full_kernel}), we replace the self-renormalization procedure with the gradient-flow matching.
\begin{equation}
\tilde{\Phi}^{\overline{\mathrm{MS}}}\left(z_1, z_2, P^z, \mu\right) 
= e^{-\delta m\tilde z}\frac{\tilde{\Phi}^R\left(z_1, z_2, P^z, a\right)}{{\cal C}_{q}(t, z_1,z_2, \mu)}.
\end{equation}

For convenience, we define the normalized matrix element
\begin{equation}\label{eq:NormM}
\hat{\Phi}^{\overline{\rm{MS}}}\left(z_1, z_2, P^z, a\right)=\frac{\tilde{\Phi}^{\overline{\rm{MS}}}\left(z_1, z_2, P^z, a\right)}{\tilde{\Phi}^{\overline{\rm{MS}}}\left(0, 0,  P^z, a\right)}.
\end{equation}

\begin{widetext}

We now specify the hybrid renormalization used in our analysis, which combines gradient flow matching with ratio renormalization. A schematic partition of the \((z_1,z_2)\) plane is shown in Fig.~\ref{fig:renorm}. For notational convenience we treat the four absolute separations
\(|z_1|,\,|z_2|,\,|z_1-z_2|\), and \(|z_1+z_2|\) as “arguments’’ that determine the region; the last of these, \(|z_1+z_2|\), is an auxiliary scale introduced only to ensure continuity of the construction near the \(z_1\!\sim\!-z_2\) diagonal (it is not a physical scale of the baryon LCDA).

The prescription proceeds case by case, always applied to the normalized matrix element \(\hat{\Phi}^{\overline{\rm MS}}\) in Eq.~(\ref{eq:NormM}). Step functions select the appropriate region and the denominator is chosen so that short-distance logarithms cancel while avoiding additional nonperturbative contamination. We adopt a single hybrid cutoff \(z_s\) satisfying \(a \ll 2z_s \ll \Lambda_{\rm QCD}^{-1}\). 

\begin{itemize}
\item \textbf{All perturbative.} 
If \(z_1\), \(z_2\), and \(z_1\!-\!z_2\) are all perturbative, corresponding to the region defined by \((|z_1|<z_s,\,|z_2|<z_s)\), \((|z_1|<z_s,\,z_s<|z_2|<2z_s)\), or \((z_s<|z_1|<2z_s,\,|z_2|<z_s)\), which is shown as the blue area in Fig.~\ref{fig:renorm}, we define

\begin{equation}
 \begin{aligned}
\hat{\Phi}^{\rm H}\left(z_1, z_2, P^z,\mu\right)=\frac{\hat{\Phi}^{\overline{\rm MS}}\left(z_1, z_2, P^z,\mu\right)}{\hat{\Phi}^{\overline{\rm MS}}\left(z_1, z_2,  0,\mu\right)} \left(\theta(2z_s-|z_1|)\theta(z_s-|z_2|)+\theta(z_s-|z_1|)\theta(|z_2|-z_s)\theta(2z_s-|z_2|)\right),
\end{aligned}
\end{equation}
with \(z_s\) as above. This ratio removes logarithms \(\ln z_1^2\), \(\ln z_2^2\), and \(\ln (z_1-z_2)^2\).

\item \textbf{ \(z_1\) perturbative.}
If \(z_1\) is perturbative while \(z_2\) and \(z_1\!-\!z_2\) are not guaranteed to be perturbative, corresponding to the region defined by \((|z_1|<z_s,\,|z_2|>2z_s)\), which is shown as the yellow vertical band in Fig.~\ref{fig:renorm}, we define

\begin{equation}
\begin{aligned}
\hat{\Phi}^{\rm H}\left(z_1, z_2, P^z,\mu\right)=\frac{\hat{\Phi}^{\overline{\rm MS}}\left(z_1, z_2, P^z, \mu\right)}{\hat{\Phi}^{\overline{\rm MS}}\left(z_1, {\rm sign}(z_2)2z_s, 0, \mu\right)}\theta(z_s-|z_1|)\theta(|z_2|-2z_s),
\end{aligned}
\end{equation}
so that \(\ln z_1^2\) cancels without injecting extra long-distance structure.

\item \textbf{ \(z_2\) perturbative.}
If \(z_2\) is perturbative while \(z_1\) and \(z_1\!-\!z_2\) are not guaranteed to be perturbative, corresponding to the region defined by \((|z_1|>2z_s,\,|z_2|<z_s)\), which is shown as the yellow horizontal band in Fig.~\ref{fig:renorm}, we define
\begin{equation}
\hat{\Phi}^{\rm H}\left(z_1, z_2, P^z,\mu\right)=\frac{\hat{\Phi}^{\overline{\rm MS}}\left(z_1, z_2, P^z, \mu\right)}{\hat{\Phi}^{\overline{\rm MS}}\left({\rm sign}(z_1)2z_s, z_2, 0, \mu\right)}\theta(|z_1|-2z_s)\theta(z_s-|z_2|).
\end{equation}

\item \textbf{\(z_1-z_2\) perturbative.}
If both \(z_1\) and \(z_2\) are not guaranteed to be perturbative while \(z_1\!-\!z_2\) is perturbative, corresponding to the region defined by \((|z_1|>z_s,\,|z_2|>z_s,\,|z_1\!-\!z_2|<z_s)\), which is shown as the yellow diagonal near \(z_1\!\sim\! z_2\) in Fig.~\ref{fig:renorm}, we define

\begin{equation}
\hat{\Phi}^{\rm H}\left(z_1, z_2, P^z,\mu\right)=\frac{\hat{\Phi}^{\overline{\rm MS}}\left(z_1, z_2, P^z, \mu\right)}{\hat{\Phi}^{\overline{\rm MS}}\left(z_1^*, z_2^*, 0, \mu\right)}\theta(|z_1|-z_s)\theta(|z_2|-z_s)\theta(z_s-|z_1-z_2|),
\end{equation}
with \(z_1^*=z_s+(z_1-z_2)\theta(z_1-z_2)\) and \(z_2^*=z_s+(z_2-z_1)\theta(z_2-z_1)\). This cancels \(\ln (z_1-z_2)^2\).

\item \textbf{\(z_1+z_2\) (auxiliary) perturbative.}
For \(|z_1|>z_s\), \(|z_2|>z_s\), and \(|z_1\!+\!z_2|<z_s\), corresponding to the region defined around \(z_1\!\sim\! -z_2\), which is shown as the blue diagonal in Fig.~\ref{fig:renorm}, we define

\begin{equation}
\hat{\Phi}^{\rm H}\left(z_1, z_2, P^z,\mu\right)= \frac{\hat{\Phi}^{\overline{\rm MS}}\left(z_1, z_2,  P^z, \mu\right)}{\hat{\Phi}^{\overline{\rm MS}}\left(z_1^{**}, z_2^{**},  0, \mu\right)}\theta(|z_1|-z_s)\theta(|z_2|-z_s)\theta(z_s-|z_1+z_2|),
\end{equation}
with \(z_1^{**}=z_s+(z_1+z_2)\theta(z_1+z_2)\) and \(z_2^{**}=-z_s+(z_2+z_1)\theta(-z_2-z_1)\). We emphasize that \(|z_1+z_2|\) is introduced only to maintain continuity of the ratio construction; it has no direct physical interpretation in the LCDA.

\item \textbf{All nonperturbative.}
Finally, if \(|z_1|>z_s\), \(|z_2|>z_s\), \(|z_1\!-\!z_2|>z_s\), and \(|z_1\!+\!z_2|>z_s\), corresponding to the gray region in Fig.~\ref{fig:renorm}, we define
\begin{equation}
\begin{aligned}
\hat{\Phi}^{\rm H}\left(z_1, z_2, P^z,\mu\right)=\frac{\hat{\Phi}^{\overline{\rm MS}}\left(z_1, z_2, P^z, \mu\right)}{\hat{\Phi}^{\overline{\rm MS}}\left({\rm sign}(z_1)z_s, {\rm sign}(z_2)2z_s, 0, \mu\right)}\theta(|z_1|-z_s)\theta(|z_2|-z_s)\theta(|z_1-z_2|-z_s)\theta(|z_1+z_2|-z_s).
\end{aligned}
\end{equation}
\end{itemize}
\end{widetext}

A detailed discussion of the hybrid renormalization scheme can be found in Refs.~\cite{Han:2023xbl,Han:2024ucv}. With the hybrid-renormalized matrix element $\hat{\Phi}^{\mathrm{H}}(z_1,z_2,P^z,\mu)$ defined above, the matching to baryon LCDAs is performed using Eq.~(\ref{lamet_fact}). The corresponding hybrid matching kernel is given in Refs.~\cite{Han:2023xbl,Han:2024ucv}.

We emphasize that, in the small flow-time limit, the flowed matching kernel ${\cal C}_{q}(t,z_1,z_2,\mu)$ becomes independent of the spatial separations. Consequently, ${\cal C}_{q}$ drops out in the definition of the hybrid-renormalized matrix elements when the small flow-time form is used. However, as discussed above, there exist regions where the hierarchy
\(
\sqrt{8t}\ll |z_1|,\ |z_2|,\ |z_1-z_2|
\)
is not satisfied. For lattice implementations it is therefore more reliable to employ the full matching kernel at finite flow time, Eq.~(\ref{full_kernel}), rather than its small flow-time approximation.

\subsection{Lattice implementation of gradient flow}
In lattice simulations one works at fixed physical flow time \(t>0\) with smearing radius \(r_{\rm sm}=\sqrt{8t}\), and the continuum limit is taken at fixed \(t\). For baryon quasi-DAs, all separations should simultaneously satisfy
\begin{equation}
\frac{1}{a} \gg \frac{1}{\sqrt{8t}} \gg \frac{1}{|z_1|},\ \frac{1}{|z_2|},\ \frac{1}{|z_1-z_2|} \gg \Lambda_{\rm QCD},
\end{equation}
so that flow-induced artifacts and discretization effects are parametrically separated from the operator scales. After the continuum extrapolation at fixed \(t\), conversion to the \(\overline{\mathrm{MS}}\) scheme can be performed using Eq.~(\ref{flow_match}) with the one-loop kernel in Eq.~(\ref{full_kernel}), evaluated at the same \(t\). To minimize logarithms in the matching coefficients, one may choose the renormalization scale such that \(\log\!\bigl(2e^{2\gamma_E}\mu^2 t\bigr)=0\).

The Wilson-line linear divergence is encoded in the factor \(e^{\delta m \tilde z}\), with \(\delta m\) given by Eq.~(\ref{self_energy}). Although this expression is derived analytically at one loop, higher-order and nonperturbative effects can modify the effective slope. Accordingly, in practice one fits the linear divergence across several flow times \(t\), using the one-loop form as a physics-motivated prior. Finally, within a window where \(t\Lambda_{\rm QCD}^2\ll 1\) and \(|z_1|,|z_2|,|z_1-z_2|\gg r_{\rm sm}\), a controlled extrapolation \(t\to 0\) can be performed.
One should note that the linear divergence is plagued by a linear renormalon~\cite{Zhang:2025jmq}, which can be treated using leading renormalon resummation~\cite{Zhang:2023bxs}.

\section{Summary and Outlook}
We presented a systematic gradient-flow based framework to extract baryon LCDAs from lattice QCD. We established a factorization relation connecting flowed Baryon quasi-DAs to their $\overline{\mathrm{MS}}$ counterparts and carried out a complete one-loop calculation beyond the small flow-time limit, obtaining the full matching kernel together with the Wilson-line linear divergence. Building on these results, we defined a hybrid renormalization scheme that combines ratio renormalization with gradient-flow matching, providing a numerically stable and theoretically controlled alternative to self-renormalization for lattice applications. Our future work will focus on obtaining lattice results using this approach.

\section*{Acknowledgement}
We thank Prof.~Wei Wang, Prof.~Nora Brambilla and Zhi-Chao Gong for valuable discussions. This work is supported in part by the National Natural Science Foundation of China under Grant No.~12125503 and 12305103. Jia-Lu Zhang is supported by T.D. Lee scholarship.

\end{document}